\documentclass[11pt]{article}

\usepackage{geometry}
\usepackage{hyperref}
\usepackage{amsmath}
\usepackage{amssymb}
\usepackage{amsthm}
\usepackage{graphicx}
\usepackage{enumitem}
\usepackage{cite}
\usepackage{url}
\usepackage{tikz}
\usetikzlibrary{positioning,fit,calc,arrows.meta}
\usepackage[most]{tcolorbox}
\usepackage{xcolor}
\definecolor{fitered}{HTML}{AE0000}
\newtcolorbox{fitobox}[1][]{%
  colback=black!5!white,
  colframe=fitered,
  coltitle=white,
  fonttitle=\bfseries,
  sharp corners,
  boxrule=0.8pt,
  #1
}

\title{\textbf{Unix Tools and the FITO Category Mistake:\\
Crash Consistency and the Protocol Nature of Persistence}}

\author{
Paul Borrill \\
DÆDÆLUS \\
\texttt{paul@daedaelus.com}
}

\date{March 2026}

\begin{document}

\maketitle

\begin{abstract}

Unix tools such as \texttt{ls}, \texttt{cp}, \texttt{mv}, and \texttt{rename}
expose a filesystem abstraction that appears to present a single,
authoritative state evolving through atomic transitions.
This abstraction is false.

We present a systematic Forward-In-Time-Only (FITO) analysis
demonstrating that the assumption of instantaneous atomic state
transitions constitutes a category mistake at every layer of the
computing stack---from ext4 journaling and delayed allocation, through
\texttt{fsync} failure semantics, NVMe Flush/FUA device behavior, and
Linux restartable sequences, down to the x86-64 CPU's own inability to
guarantee atomic supervisor entry under Non-Maskable Interrupts.

We prove a formal impossibility result: no syscall-based persistence
primitive can define a commit boundary under failure, because the
syscall return value is consistent with multiple materially different
persistence states across Linux filesystems.
We identify cross-layer temporal assumption leakage as the structural
mechanism by which the category mistake propagates, and show that the
entire storage stack forms a recursive chain of non-atomic dependencies
whose apparent atomicity reflects mathematical impossibility
(Herlihy, 1991), not merely engineering deficiency.

An appendix documents the real-world consequences: cascading cloud
outages at Google, AWS, Meta, and Cloudflare driven by retry
amplification; database corruption from fsync failures in PostgreSQL,
etcd, and MySQL; silent data corruption at CERN, NetApp, and Meta;
AI training waste consuming 12--43\% of compute budgets at scale; and
financial system failures totaling billions of dollars annually.
These consequences trace to a single structural cause: systems designed
around the FITO assumption, compensating for its failure with
retry-and-recover protocols that amplify the very failures they
attempt to mask.

\end{abstract}

\section{Introduction}

Unix presents a deceptively simple abstraction: a filesystem consisting of
named objects, manipulated by commands that appear to execute atomically.

This abstraction implies the existence of a single authoritative state that
evolves through a sequence of discrete transitions.

However, the underlying implementation is fundamentally protocol-based.

Filesystem operations are implemented through asynchronous writeback,
journaling, and crash recovery mechanisms.

The apparent instantaneous transition observed by applications is merely the
visible endpoint of a protocol.

This mismatch between abstraction and implementation constitutes a Category
Mistake.

\section{Crash Consistency and Sequential Illusion}

Application developers commonly assume sequential crash consistency:

\begin{quote}
If operations A, B, C are executed in order, then after a crash, the system
reflects either A, A+B, or A+B+C.
\end{quote}

This assumption is false in ext4 and most modern filesystems.

Ext4 implements weaker crash consistency semantics, allowing reordering of
writes and metadata operations unless explicit persistence barriers such as
\texttt{fsync()} are used \cite{ferrite}.

Thus:

\begin{itemize}
\item Execution order does not imply persistence order.
\item Namespace atomicity does not imply data durability.
\item Completion of a system call does not imply persistence.
\end{itemize}

This is the filesystem-level manifestation of FITO.

\section{ext4 Architecture and Journaling Semantics}

ext4 implements journaling using the JBD2 journal layer.

Three journaling modes exist \cite{ext4docs}:

\begin{enumerate}
\item data=journal
\item data=ordered
\item data=writeback
\end{enumerate}

Only metadata is guaranteed to be journaled in ordered and writeback modes.

File data may be written later.

Thus metadata indicating a file exists may be persisted before the file
contents themselves are durable.

\subsection{Delayed Allocation}

ext4 implements delayed allocation to improve performance.

Blocks are not allocated at write time but later during writeback.

This introduces temporal separation between:

\begin{itemize}
\item logical write completion
\item physical block allocation
\item persistence
\end{itemize}

This separation invalidates the assumption that write completion defines a
temporal commit point.

\section{The Rename Operation: Atomic Namespace, Non-Atomic Persistence}

POSIX defines rename as an atomic namespace operation \cite{posix_rename}.

However:

\begin{itemize}
\item rename does not guarantee durability
\item rename does not guarantee ordering relative to data writes
\end{itemize}

Correct durable replacement requires:

\begin{enumerate}
\item write temp file
\item fsync(temp)
\item rename(temp, target)
\item fsync(parent directory)
\end{enumerate}

Without directory fsync, the rename itself may be lost.

\subsection{The Write-Sync-Rename Pattern}

Real-world systems routinely depend on this four-step protocol.
Git, for example, implements its reference updates through lock files
created with \texttt{open(O\_CREAT|O\_EXCL)} and committed via the
\texttt{write} $\to$ \texttt{fsync} $\to$ \texttt{rename} sequence
\cite{borrill2026icloud}.

This pattern is widely treated as providing transactional semantics:
either the update completes fully, or the repository is unchanged.
But as the preceding analysis shows, the transactional guarantee
is contingent on every layer of the persistence stack---from
\texttt{fsync} through the block layer to NVMe Flush---behaving
correctly. When cloud-synchronized storage systems such as iCloud
interpose an eventually consistent replication layer beneath this
protocol, the lock files themselves become subjects of asynchronous
propagation, defeating the very mutual exclusion they were designed
to provide.

Thus rename is atomic in namespace semantics but not in persistence semantics.

\section{ext4 Fast Commit}

ext4 Fast Commit was introduced to improve performance of workloads that rely
heavily on fsync for correctness \cite{fastcommit}.

Fast Commit logs minimal metadata sufficient to replay recent operations.

This feature exists precisely because applications require explicit protocol
barriers to obtain correct semantics.

The existence of Fast Commit confirms that correctness depends on protocol
execution rather than instantaneous state transitions.

\section{Formal FITO Analysis of Unix Tools}

We now analyze canonical Unix tools.

\subsection{ls}

Assumes existence of coherent directory state.

Reality: produces observational projection of evolving namespace.

\subsection{cp}

Assumes atomic duplication.

Reality: multi-step protocol with partial externally visible states.

\subsection{mv}

Assumes atomic move.

Reality:

\begin{itemize}
\item atomic within filesystem
\item copy-delete protocol across filesystems
\end{itemize}

\subsection{grep}

Assumes stable file contents.

Reality: file may change during read.

\subsection{rm}

Assumes immediate deletion.

Reality: removes namespace binding, not underlying object immediately.

\section{VERITAS File System}

The VERITAS File System (VxFS) implements intent logging for crash recovery
\cite{vxfsintent}.

Intent logging records metadata operations in a log.

After crash, operations are replayed.

This mechanism preserves filesystem structural consistency but does not alter
the protocol nature of filesystem operations.

It confirms that filesystem evolution is log-based rather than instantaneous.

\subsection{Practitioner Experience}

Direct experience with enterprise filesystem engineering at VERITAS
exposed the gap between documented semantics and deployed behavior.
The author participated in extended discussions with VERITAS filesystem
engineers regarding the misuse of \texttt{fsync} as a durability
primitive, arguing that the atomicity violations observed in storage
hardware made the common usage patterns unreliable
\cite{mulliganstew2024}.

The core insight emerging from those discussions---and later formalized
in the Mulligan Stew discussion group---is that the correct
specification of an ``atomic'' operation is not instantaneous state
transition but rather:

\begin{fitobox}[title=Completeness Requirement]
Either all constituent operations happen in the correct sequence and
complete, or all constituent operations are reversed in the correct
reverse sequence and complete.
\end{fitobox}

This formulation replaces the FITO assumption of a single commit instant
with a protocol that must converge in both the forward and reverse
directions. It directly anticipates the reversible subtransaction model
discussed in Section~\ref{sec:atomicity-stack} and establishes that the
problems identified in this paper were recognized in industrial practice
well before they were formally characterized.

\section{VERITAS Volume Manager}

VERITAS Volume Manager implements Dirty Region Logging (DRL) to track modified
regions in mirrored volumes \cite{vxvmdrl}.

DRL enables efficient recovery by identifying regions requiring resynchronization.

This explicitly acknowledges that storage evolution occurs over time and must
be reconstructed through protocol analysis.

\section{Fundamental Category Mistake}

Unix presents filesystem operations as state transitions.

Implementation reality:

Filesystem operations are protocol executions over time.

Thus filesystem semantics are fundamentally causal and protocol-based, not
instantaneous state-based.

This mismatch constitutes a Category Mistake.

\section{Cross-Layer Temporal Assumption Leakage}

The FITO category mistake is not confined to individual system calls.
It manifests wherever one abstraction layer assumes stronger temporal
guarantees than the layer below it actually provides.

This pattern---\emph{cross-layer temporal assumption leakage}---recurs
throughout the storage stack:

\begin{enumerate}
\item Applications assume \texttt{fsync} defines a durability boundary,
      but the filesystem may mark pages clean without achieving
      persistence (Section~\ref{sec:fsync-failure}).
\item File systems assume the block layer preserves ordering, but
      NVMe Flush may be a no-op.
\item Build systems assume atomic visibility of dependencies, but
      cloud-synchronized storage provides only eventual consistency.
\item Backup systems (e.g., Time Machine) capture snapshots of local
      state that may not reflect the cloud-synchronized global state.
\end{enumerate}

Prior work on deterministic document builds \cite{borrill2026hermetic}
demonstrates this leakage in a minimal, reproducible form: LaTeX
compilation implicitly assumes a stable, atomically visible dependency
set during the build process---an assumption that holds on a local
filesystem but is violated by cloud-synchronized storage.
The resulting failures are not compiler bugs but structural mismatches
between the temporal model assumed by the application and the weaker
model provided by the storage substrate.

The generalization is that any system operating above an eventually
consistent or asynchronously replicated storage layer inherits the FITO
category mistake \emph{unless} it explicitly accounts for the protocol
nature of the layers below.

\section{Failure of fsync as a Temporal Commit Boundary}
\label{sec:fsync-failure}

The fsync system call is widely assumed to define a commit barrier, establishing
durable persistence of file data.

Recent research demonstrates that this assumption is false.

\subsection{Undefined State After fsync Failure}

File systems may mark pages clean after fsync failure even when data is not
durable \cite{rebello2021}.

This breaks the invariant:

\[
clean(page) \Rightarrow durable(page)
\]

Applications cannot determine the true persistence state.

\subsection{Structural Inconsistency}

ext4 may enter structurally inconsistent states after fsync failure, such as
inodes referencing blocks not marked allocated \cite{rebello2021}.

The filesystem may incorrectly report itself clean, preventing repair.

\subsection{Temporal Dislocation of Error Reporting}

fsync failure may not be reported at the time of the failed write, but at a
subsequent unrelated operation \cite{rebello2021}.

This breaks causal temporal reasoning.

\subsection{Non-Convergent Persistence}

Subsequent fsync calls do not retry failed writes.

Thus fsync does not guarantee eventual convergence to a consistent durable state.

\subsection{Impossibility of a Syscall-Observable Persistence Boundary Under fsync Failure}

We now formalize the failure of \texttt{fsync} to define a persistence
boundary in terms of causal protocol convergence.

\paragraph{Definition 1 (Persistence protocol).}
Let the persistence protocol be a distributed state machine consisting
of causal layers:
\[
N =
\{
Application,\;
PageCache,\;
FilesystemJournal,\;
BlockLayer,\;
ControllerCache,\;
PersistentMedia
\}
\]
Each layer $i \in N$ has state $S_i(t)$.
Define durability as convergence:
\[
durable \iff \forall i \in N,\; S_i \in Committed
\]
This definition captures persistence as a convergence property of the
distributed protocol.

\paragraph{Definition 2 (Commit boundary).}
A syscall $X$ defines a commit boundary iff its return value uniquely
determines whether the durability condition holds:
\[
Return(X) \Rightarrow durable \;\; \text{or} \;\; \neg\, durable
\]
and the application can safely proceed based solely on this observation.

\paragraph{Lemma (Impossibility of fsync Commit Boundary Under Failure).}
Under observed Linux filesystem semantics, \texttt{fsync} does not
define a commit boundary when it fails.

\paragraph{Proof.}
Rebello et al.\ demonstrate that after fsync failure, multiple Linux
filesystems mark affected pages clean even when they have not been
written to durable media \cite{rebello2021}.
Specifically, ext4, XFS, and Btrfs clear the page dirty bit prior to
I/O submission (via \texttt{clear\_page\_dirty\_for\_io}) and do not
restore it on failure, leaving pages marked clean despite incomplete
persistence.

Thus the syscall return value
\[
Return(fsync) = failure
\]
is consistent with multiple materially different protocol states:
\begin{gather*}
\exists \Sigma_1, \Sigma_2 :\;
Return(fsync)=failure \\
\land\;
durable(\Sigma_1) \neq durable(\Sigma_2)
\end{gather*}
where $\Sigma_1$ and $\Sigma_2$ differ in persistence convergence.
(For example, ext4 and XFS retain the latest write in memory, while
Btrfs reverts in-memory page contents to the previous on-disk state.)

Therefore syscall observation does not uniquely determine durability.
Hence fsync failure does not define a commit boundary. \qed

\paragraph{Corollary 1 (Retry Non-Soundness).}
Application retry based solely on fsync failure is not a correct
recovery strategy, because subsequent fsync calls do not necessarily
retry failed writes and the protocol may already have transitioned to
a state treating data as clean \cite{rebello2021}.
(XFS retries metadata writes by default, but this does not extend to
general data page recovery on ext4 or Btrfs.)

\paragraph{Corollary 2 (Epistemic Ambiguity).}
After fsync failure, application-visible state does not determine
protocol convergence.
Durability becomes epistemically unknowable at the syscall boundary.

\paragraph{Corollary 3 (FITO Category Mistake).}
The assumption that fsync defines a temporal persistence boundary
commits a Forward-In-Time-Only category mistake.
fsync is an attempt to drive a distributed persistence protocol toward
convergence, not an instantaneous state transition.

\paragraph{Theorem (Persistence Cannot Be Observed at the Syscall Boundary).}
There exists no syscall-based persistence primitive whose return value
alone determines durability under failure conditions in the Linux
persistence stack.

\paragraph{Proof.}
Follows directly from the lemma: if fsync failure does not uniquely
determine convergence state, no syscall return value can serve as a
persistence boundary. \qed

\begin{fitobox}[title=Key Insight]
Durability is not a property of an event in time.
It is a property of a distributed protocol reaching a fixed point.
\end{fitobox}

\section{NVMe Flush, FUA, and Power-Loss Protection}

At the lowest layer, file-system durability depends on the persistence semantics
of the storage device.

For NVMe devices, the specification exposes two relevant mechanisms:
(i) an explicit \emph{Flush} command, and (ii) a \emph{Force Unit Access} (FUA)
bit on Write commands.

Linux explicitly models controls for writeback cache behavior in the block
layer \cite{linux_writeback_cache_control}.
Consequently, any claim of durability obtained via \texttt{fsync()} is an
implicit claim about the behavior of Flush and FUA at the storage device.

\subsection{Flush as a Conditional Durability Mechanism}

NVMe Flush is defined to force data from a \emph{volatile write cache} to
non-volatile media.

However, the specification also notes that if a volatile write cache is not
present or not enabled, Flush commands complete successfully and have no effect.
Thus, ``Flush succeeded'' does not uniquely identify a durability boundary; it
may be a no-op depending on device configuration and cache design
\cite{nvme_flush_noop}.

\subsection{FUA and the Semantics of Write Completion}

NVMe further distinguishes write completion from durable persistence by defining
an explicit exception: if the controller supports a volatile write cache, the
cache is enabled, the FUA bit is not set, and no Flush has successfully completed
for the namespace, then later reads may return older data rather than the data
from the most recently completed write \cite{nvme_cmdset_2024}.

This is a direct specification-level rejection of the naive FITO assumption
``write completed implies durable.''

\subsection{Linux Mapping: Flush and FUA as Block-Layer Primitives}

The Linux block layer exposes two mechanisms to control caching behavior at the
device boundary: a forced cache flush and an FUA flag for requests
\cite{linux_queue_fua}.
File systems and applications that rely on
\texttt{fsync()} are therefore implicitly relying on the correctness of this
translation from file API semantics into NVMe Flush and Write(FUA) semantics.

\subsection{Power-Loss Protection Reclassifies Cache Volatility}

A subtle but decisive case arises when a device contains a write cache but can
guarantee that cached data is written to non-volatile media on loss of power.
In this case, the cache is considered \emph{non-volatile} and the volatile write
cache feature does not apply; configuration controls for ``volatile write cache''
may therefore have no effect \cite{ms_nvme_vwc_note,ssd_plp}.

Thus, the same command interface may correspond to materially different causal
realities: a volatile cache requiring Flush/FUA to establish a durability
boundary, or a non-volatile (PLP-backed) cache whose persistence behavior is
fundamentally different.

\subsection{Verification: What the System Can and Cannot Prove}

To reason correctly about durability, systems must distinguish at least four
questions:

\begin{enumerate}[label=(Q\arabic*)]
\item Is a volatile write cache present (Identify Controller VWC field)?
\item Is volatile write caching enabled (Feature Identifier 06h)?
\item Does the OS/device stack support FUA for writes (e.g., Linux queue FUA support)?
\item Under injected failure, does a reported fsync failure leave the file system
      in a well-defined epistemic state?
\end{enumerate}

The final question is not theoretical: recent work shows that after fsync fails,
multiple Linux file systems mark pages clean, rendering application-level retry
ineffective and leaving application developers uncertain about file-system state
\cite{rebello2021}.

\subsection{FITO Interpretation}

The durability boundary is not a temporal instant.

It is an emergent protocol state spanning the application, file system, Linux
block layer, NVMe controller firmware, cache design, and media.

The persistence boundary depends on (a) whether the device has a volatile cache,
(b) whether that cache is enabled, (c) whether FUA is set on relevant writes, and
(d) whether a Flush has completed.

Thus, even at the device layer, durability cannot be modeled as an instantaneous
event in a single linear timeline; it is an emergent property of a multi-layer
protocol spanning the OS, block layer, controller firmware, and media.

Treating durability as a single linear ``commit time'' constitutes a
Forward-In-Time-Only category mistake.

\section{CPU-Level Atomicity: Restartable Sequences and NMI}

The preceding sections demonstrate that atomicity fails at every layer
of the storage stack from application through NVMe media. We now show
that the failure extends even deeper: the CPU itself cannot guarantee
atomicity of instruction sequences under interrupts.

\subsection{Non-Maskable Interrupts and Supervisor Entry}

On x86-64, the transition from user mode to supervisor mode via
\texttt{SYSCALL} requires the kernel to execute \texttt{SWAPGS} (to
load the kernel's GS base) and load a kernel stack pointer before it
can safely handle any exception. These operations are not atomic with
respect to Non-Maskable Interrupts (NMIs).

Torvalds identified this as a fundamental architectural deficiency:
NMIs can fire during the \texttt{SWAPGS}/stack-switch window, creating
a state in which the processor is neither fully in user mode nor fully
in kernel mode \cite{torvalds_nmi}.

This is not a software bug but a hardware design limitation: the x86-64
\texttt{SYSCALL} instruction was designed without considering the
interaction between supervisor entry and asynchronous interrupts. Both
Intel (with the FRED---Flexible Return and Event Delivery---mechanism)
and AMD (with Supervisor Entry Extensions) have proposed hardware
modifications to close this gap, confirming that it is a recognized
architectural deficiency rather than a theoretical concern
\cite{fred_intel}.

\subsection{Linux Restartable Sequences}

The Linux kernel's response to the impossibility of cheap user-space
atomicity is \emph{restartable sequences} (rseq), merged in Linux 4.18
after a five-year development effort \cite{desnoyers_rseq}.

The rseq mechanism works as follows:

\begin{enumerate}
\item A thread registers a per-thread \texttt{struct rseq} with the
      kernel via the \texttt{rseq(2)} system call.
\item Before entering a critical section, the thread sets a pointer to
      metadata describing the section's instruction range and abort
      handler.
\item The thread executes a preparatory sequence followed by a single
      atomic commit instruction.
\item If the kernel preempts the thread, delivers a signal, or migrates
      it to another CPU during the preparatory phase, it redirects
      execution to the abort handler, which typically retries.
\end{enumerate}

The critical insight is that rseq does not \emph{prevent} interruption;
it \emph{detects and recovers from} interruption by restarting the
sequence.

\begin{fitobox}[title=CPU-Level Insight]
Atomicity is not a property of the execution but an emergent
outcome of a retry protocol---precisely the pattern identified
at every layer of the stack, from database transactions through
NVMe firmware to CPU instruction sequences.
\end{fitobox}

\subsection{FITO Interpretation}

Restartable sequences confirm that even at the CPU instruction level,
atomicity is not an instantaneous property but a protocol outcome.

The ``atomic'' per-CPU operation is actually:

\begin{enumerate}
\item Attempt the critical section.
\item If interrupted, abort and retry.
\item Repeat until the section completes without interruption.
\end{enumerate}

This is a protocol execution over time, not an instantaneous state
transition. The rseq mechanism achieves the \emph{appearance} of
atomicity through exactly the same pattern as filesystem journaling,
fsync barriers, and NVMe Flush commands: a multi-step protocol that
converges to a consistent state only when no failures intervene.

The x86-64 NMI problem and the rseq mechanism together establish that
the FITO category mistake is not confined to software abstractions. It
is present at every layer of the system, from application semantics
through the CPU's own interrupt architecture. Treating any layer as
providing instantaneous atomic transitions constitutes a category
mistake.

\section{Formal Causal Model of Persistence}

We now formalize persistence as a causal convergence property rather than a temporal event.

% ============================================================
% Figure 1: Causal Persistence Graph (Protocol Convergence)
% ============================================================
\begin{figure}[htbp]
\centering
\begin{tikzpicture}[
  font=\footnotesize,
  node distance=6mm and 8mm,
  layer/.style={
    draw,
    rounded corners,
    align=center,
    inner sep=4pt,
    text width=38mm
  },
  box/.style={
    draw,
    rounded corners,
    align=left,
    inner sep=4pt,
    text width=42mm,
    font=\scriptsize
  },
  fail/.style={draw, circle, inner sep=1.5pt, fill=white},
  arrow/.style={->, thick}
]
% --- Nodes (top to bottom) ---
\node[layer] (app) {Application\\(update protocol)};
\node[layer, below=of app] (vfs) {VFS \& Syscalls\\(\texttt{write}, \texttt{rename}, \texttt{fsync})};
\node[layer, below=of vfs] (pc) {Kernel Page Cache\\(dirty pages)};
\node[layer, below=of pc] (fs) {Filesystem Layer\\(ext4, XFS, Btrfs)};
\node[layer, below=of fs] (jbd) {Journal / Log\\(JBD2 / intent log)};
\node[layer, below=of jbd] (blk) {Linux Block Layer\\(queues, barriers)};
\node[layer, below=of blk] (nvme) {NVMe Controller\\(cache, FUA, Flush)};
\node[layer, below=of nvme] (media) {Persistent Media\\(NAND / disk)};
% --- Main causal arrows ---
\draw[arrow] (app) -- (vfs);
\draw[arrow] (vfs) -- (pc);
\draw[arrow] (pc) -- (fs);
\draw[arrow] (fs) -- (jbd);
\draw[arrow] (jbd) -- (blk);
\draw[arrow] (blk) -- (nvme);
\draw[arrow] (nvme) -- (media);
% --- Failure injection points (right side) ---
\node[fail, right=12mm of pc] (f1) {};
\node[fail, right=12mm of jbd] (f2) {};
\node[fail, right=12mm of blk] (f3) {};
\node[fail, right=12mm of nvme] (f4) {};
\draw[arrow] (pc.east) -- (f1.west);
\draw[arrow] (jbd.east) -- (f2.west);
\draw[arrow] (blk.east) -- (f3.west);
\draw[arrow] (nvme.east) -- (f4.west);
\node[box, right=3mm of f1] (t1) {F1: page marked clean\\after failed \texttt{fsync}};
\node[box, right=3mm of f2] (t2) {F2: journal-block write\\failure $\Rightarrow$ abort/RO};
\node[box, right=3mm of f3] (t3) {F3: missing/failed\\flush barrier};
\node[box, right=3mm of f4] (t4) {F4: volatile cache,\\no FUA / no Flush};
% --- PLP callout (below media, right-aligned) ---
\node[box, right=12mm of media, yshift=-2mm] (plp)
  {Power-Loss Protection (PLP)\\reclassifies cache as non-volatile};
\draw[arrow] (plp.north) -- ++(0,4mm) -| (nvme.south east);
% --- rename callout (left side) ---
\node[box, left=12mm of fs] (ren)
  {\texttt{rename()} is namespace-atomic\\but not necessarily durable\\without directory \texttt{fsync}};
\draw[arrow] (ren.east) -- (fs.west);
\end{tikzpicture}
\caption{Causal persistence graph: durability is not a temporal instant but an
emergent protocol state spanning application logic, the kernel,
filesystem/journal, the Linux block layer, the NVMe controller
(Flush/FUA/cache), and persistent media.
Failure points (F1--F4) indicate where a linear ``commit-time'' mental model
breaks.}
\label{fig:causal-persistence-graph}
\end{figure}
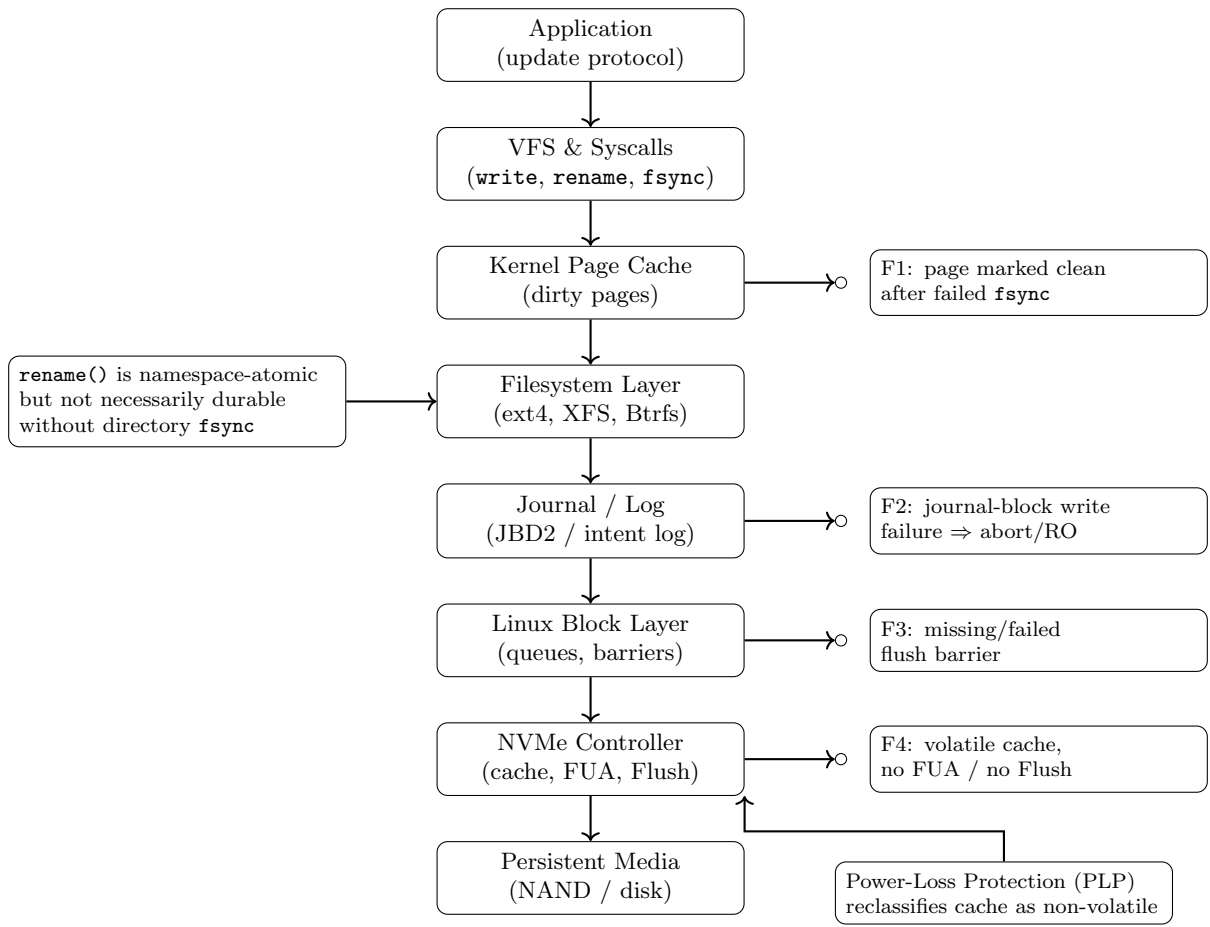

Let the persistence protocol consist of nodes:

\[
N =
\{
Application,
KernelPageCache,
FilesystemJournal,
BlockLayer,
ControllerCache,
PersistentMedia
\}
\]

Each node has state:

\[
S_i(t)
\]

Persistence is defined as convergence:

\[
durable \iff \forall i \in N,\; S_i = committed
\]

No single temporal instant guarantees this property.

Instead, persistence is achieved only when the distributed protocol converges.

\subsection{Non-Existence of a Temporal Commit Boundary}

Suppose a commit boundary exists at time $t_c$.

Then:

\[
\exists t_c \quad \forall t \ge t_c,\; durable(t)
\]

However, fsync failure results demonstrate that filesystem and device layers may diverge after fsync returns.

Thus:

\[
\nexists t_c
\]

\begin{fitobox}[title=Formal Result]
Persistence is not a temporal event.
It is a protocol convergence property.
\end{fitobox}

\subsection{Verification Impossibility}

Applications observe only syscall return values:

\[
observe = syscall\_return
\]

But durability depends on hidden device state:

\[
durable = f(hidden\_protocol\_state)
\]

Therefore:

\[
observe \not\Rightarrow durable
\]

Durability cannot be verified from system calls alone.

\subsection{FITO Interpretation}

The Unix abstraction projects protocol convergence onto a linear time axis.

This projection is incorrect.

Durability exists only as a property of causal protocol convergence.

Treating persistence as a temporal event constitutes a Forward-In-Time-Only category mistake.

\section{The Atomicity Stack: A Chain of Non-Atomic Dependencies}
\label{sec:atomicity-stack}

The preceding analysis has traced the FITO category mistake through
individual layers: filesystem semantics, journaling, fsync, and NVMe
device behavior. We now observe that these layers form a dependency
chain in which each layer's atomicity guarantee depends on the layer
below it:

\begin{enumerate}
\item The database relies on the transaction log.
\item The transaction log relies on the filesystem.
\item The filesystem relies on \texttt{fsync}.
\item \texttt{fsync} relies on storage hardware (NVMe Flush/FUA).
\item Storage hardware relies on the CPU executing instructions atomically.
\item The CPU itself cannot guarantee atomicity under NMI
      \cite{torvalds_nmi}.
\end{enumerate}

No layer in this chain provides unconditional atomicity.
Each layer assumes the layer below it will behave correctly, but as
demonstrated in the preceding sections, every layer violates this
assumption under failure.

This structure constitutes a \emph{recursive} category mistake:
the apparent atomicity at each level is not a property of that level
but an assumption about the level below. When any layer fails, the
recovery model reduces to what practitioners call ``smash and
restart''---discard state and rebuild from the most recent known-good
checkpoint.

\subsection{Relationship to the Impossibility of Atomicity}

This layered failure mode is not merely an engineering deficiency.
Herlihy's consensus hierarchy establishes that read/write registers
have consensus number 1: they cannot solve consensus between even two
processes \cite{herlihy1991}.

Since filesystem operations ultimately reduce to reads and writes
against storage media, no amount of protocol layering can conjure
atomicity from primitives that are mathematically incapable of
providing it \cite{borrill2026impossibility}.

The Unix tool abstraction thus commits two category mistakes
simultaneously:

\begin{enumerate}
\item Treating protocol executions as instantaneous state transitions
      (the temporal FITO mistake).
\item Treating compositions of non-atomic primitives as atomic
      operations (the algebraic impossibility mistake).
\end{enumerate}

\subsection{Implications for Recovery}

The standard recovery model---write-ahead logging with redo/undo---is
itself a FITO construction: it assumes a linear log that can be
replayed forward from a known-good state. But as the fsync failure
results demonstrate \cite{rebello2021}, the ``known-good state'' may
itself be corrupted, and the log replay may not converge.

An alternative approach, proposed in prior work on reversible
subtransactions \cite{borrill2026rethinking}, rejects the FITO recovery
model entirely. Instead of assuming a linear log that replays forward,
reversible protocols construct inverse operations that can unwind
partial state changes without requiring a global checkpoint.

This approach treats atomicity not as an assumption but as a
constraint to be enforced through protocol design---specifically,
through primitives with infinite consensus number such as
memory-to-memory swap \cite{herlihy1991,borrill2026impossibility}.

\section{Conclusion}

ext4 and enterprise filesystems such as VxFS and VxVM demonstrate that filesystem
operations are protocol executions involving journaling, delayed allocation,
and crash recovery.

The apparent instantaneous state transitions observed by applications are
projections of underlying protocol executions.

Treating these operations as atomic temporal events constitutes a FITO category
mistake.

Moreover, the entire storage stack---from database transactions through
filesystem journals, \texttt{fsync}, and NVMe device semantics---forms a chain
of non-atomic dependencies. Each layer assumes atomicity from the layer below,
yet no layer provides it unconditionally. This recursive dependence on
non-atomic primitives is not merely an implementation gap; it reflects the
mathematical impossibility of achieving consensus through reads and writes
alone \cite{herlihy1991}.

Correct reasoning requires recognizing filesystem operations as protocol
executions over causal structures rather than instantaneous state transitions.
Where atomicity is required, it must be constructed from primitives that are
algebraically capable of providing it---such as memory-to-memory swap---rather
than assumed from compositions of reads and writes that cannot.

\appendix

\section{Real-World Consequences of the FITO Category Mistake}
\label{app:consequences}

The preceding analysis demonstrates that the FITO category mistake
pervades the storage stack from application semantics through CPU
architecture. This appendix documents the real-world consequences:
cascading outages, data corruption, and economic losses that arise
directly from treating protocol executions as instantaneous state
transitions.

\subsection{Cloud Infrastructure: Retry Amplification at Scale}

When FITO systems encounter uncertainty, they interpret it as failure
and inject retries. Under load, retries increase load, which increases
uncertainty, which increases retries---a positive-feedback loop that
transforms localized faults into planetary-scale outages.

\paragraph{Google Cloud, June 12, 2025.}
A null pointer exception in Google Cloud's Service Control system---the
authentication gatekeeper for all GCP API calls---cascaded through
Spanner's global replication into a 7-hour outage affecting 70+
services across 40+ regions, generating 1.4 million user reports.
Cloudflare, OpenAI, Spotify, Discord, and Snapchat were collateral
casualties. The root cause was not the null pointer but the ``herd
effect'': Service Control tasks restarted en masse, flooding regional
Spanner databases without backoff and creating a secondary cascade
that extended the outage for hours \cite{gcp_june2025}.

\paragraph{AWS DynamoDB, October 19--20, 2025.}
A race condition in DynamoDB's internal DNS management microservice
caused a 15-hour outage across 60+ countries. When DNS was restored
and DynamoDB reconnected, millions of EC2 instances and Lambda
functions simultaneously retried failed connections, creating a
thundering herd that immediately re-overwhelmed the database control
plane and caused DNS to fail again. AWS ultimately disabled DynamoDB
DNS automation worldwide and added manual review---confirming that
retry amplification is architectural, not implementational.
Parametrix estimated direct financial losses of \$500--650 million
for US companies alone \cite{aws_oct2025}.

\paragraph{AWS Kinesis, November 25, 2020.}
Adding capacity to Kinesis servers exceeded the operating system's
maximum thread limit, corrupting the backend shard routing map.
Because many AWS services depend on Kinesis for internal communication,
the failure cascaded across CloudWatch, Cognito, Lambda, IoT Core,
and EventBridge. Negative DNS lookups persisted in caches, and client
retries extended the outage to 17 hours \cite{aws_kinesis2020}.

\paragraph{Meta, October 4, 2021.}
A routine maintenance command intended to assess backbone capacity
accidentally disconnected all Facebook data centers from the internet.
DNS servers lost connectivity, declared themselves unhealthy, and
withdrew their BGP advertisements. The resulting retry storm from
3.5 billion users' devices produced a 40\% increase in global DNS
traffic, increasing latency for unrelated web applications worldwide.
Recovery required physical data center access because internal
operations tools depended on the downed infrastructure. Duration:
6--7 hours; estimated advertising revenue loss: \$60+ million
\cite{meta_oct2021}.

\paragraph{Cloudflare, September 12, 2025.}
A React \texttt{useEffect} bug in the Cloudflare dashboard created
an object in the dependency array that was recreated on every state
change, generating excessive API calls. The calls overwhelmed the
Tenant Service during a concurrent update, cascading to all dashboard
APIs. Cloudflare's post-mortem explicitly identified the need for
randomized retry delays to prevent thundering herd effects
\cite{cloudflare_sept2025}.

\subsection{Database Corruption: fsync Failures in Practice}

The theoretical fsync failure modes analyzed in
Section~\ref{sec:fsync-failure} manifest as real data loss in
production systems.

\paragraph{PostgreSQL ``fsyncgate'' (2018--2019).}
Craig Ringer reported to the PostgreSQL mailing list that
PostgreSQL's fsync error handling was fundamentally broken on Linux.
When buffered writes failed, the kernel marked pages with an error
flag (\texttt{AS\_EIO}) but cleared the flag after the first failed
fsync. On retry, fsync returned success because the error flag was
already cleared---even though data never reached disk. The fix,
merged in PostgreSQL 12, was to \emph{panic} on fsync failure
rather than retry, because no correct retry strategy exists under
the observed kernel semantics \cite{danluu_fsyncgate}.

This is a direct empirical confirmation of our Lemma:
application-visible observation does not determine persistence state.

\paragraph{etcd Raft log corruption (2021).}
etcd versions 3.5.0--3.5.2 experienced corrupted Raft logs when
cluster leaders were killed under high load, with committed
transactions not reflected on all members after restart. The
root cause was fsync performance bottlenecks: slow fsync on the
Raft log caused leader unavailability because the leader could
not commit new proposals while waiting for disk
\cite{etcd_corruption}.

\paragraph{MySQL InnoDB doublewrite buffer.}
The InnoDB doublewrite buffer exists solely because 16\,KB InnoDB
pages can be torn during a crash that occurs mid-write, leaving
pages with a mix of old and new data. This mechanism has been
present since MySQL 5.1 and imposes measurable write amplification
on every InnoDB deployment worldwide---a permanent performance tax
to compensate for the non-atomicity of page writes
\cite{mysql_doublewrite}.

\paragraph{NVMe power loss data loss.}
Independent testing (2022) demonstrated that consumer NVMe SSDs
from SK Hynix and Sabrent lost data after power interruption despite
completed fsync calls, while Samsung and Western Digital drives
retained data. The difference: hardware Power-Loss Protection (PLP)
capacitors. Enterprise SSDs with PLP cost significantly more,
representing a direct economic consequence of the non-atomic
persistence stack \cite{nvme_plp_test}.

\subsection{Silent Data Corruption}

\paragraph{CERN (2007).}
A six-month study writing approximately 97 petabytes across 3,000
nodes found 128\,MB of permanently corrupted data---a rate of
$1.2 \times 10^{-9}$---with 500 errors detected on 100 nodes
\cite{cern_corruption}.

\paragraph{NetApp (2008).}
A 41-month study across 1.53 million disk drives documented over
400,000 checksum mismatch instances, of which 30,000 were not
detected by the RAID controller and were found only during
scrubbing. Nearline disks corrupted at 10$\times$ the rate
of enterprise drives \cite{netapp_corruption}.

\paragraph{Meta silent data corruption (2021--2026).}
Meta's fleet analysis found approximately 1 in 1,000 machines
affected by silent data corruption (SDC). For large-scale AI
training runs, SDC events are expected every one to two weeks.
The problem has worsened with process shrinks: soft error rates
increased from one failure per year at 65\,nm to one failure per
1.5 hours at 16\,nm \cite{meta_sdc,ocp_sdc}.

\subsection{AI Training Infrastructure}

The FITO category mistake imposes enormous costs on AI training,
where checkpoint-and-retry is the dominant fault tolerance
strategy.

\paragraph{Failure rates at scale.}
Meta's Llama~3 training on 16,384 NVIDIA H100 GPUs experienced 419
unexpected interruptions over 54 days---one failure every 3 hours on
average. GPU and NVLink failures accounted for 30\%, HBM3 memory
failures 17\%, and software/network issues 41\% \cite{llama3_failures}.
Academic analysis of Meta's production clusters projects that a
131,072-GPU job would experience mean time to failure of approximately
14 minutes \cite{meta_mttf_study}.

\paragraph{Checkpoint overhead.}
Checkpoint overhead consumes 12--43\% of total training time at
scale. A 16,000-accelerator cluster requires approximately 155
checkpoints per day (one every 9.3 minutes); at 100,000
accelerators, this rises to 967 checkpoints per day (one every
1.5 minutes). For frontier model training runs approaching
\$1 billion in compute cost, this overhead represents
\$120--430 million in pure waste \cite{google_checkpoint,aws_checkpoint}.

\paragraph{Recovery time.}
Traditional checkpoint-based recovery requires 15--30+ minutes per
failure event. At one failure every 3 hours, a 16,384-GPU cluster
loses 8--17\% of its training time to recovery alone---on top of
the checkpoint overhead.

\subsection{Financial Systems and Critical Infrastructure}

\paragraph{Knight Capital, August 1, 2012.}
A deployment script that failed silently when SSH connections dropped
left one of eight trading servers running deprecated code. The
desynchronization between order initiation and completion recording
caused the server to transmit 4+ million unintended orders in 45
minutes, producing \$460+ million in losses. The SEC subsequently
charged Knight Capital with market access rule violations
\cite{knight_capital}.

\paragraph{CME Globex, November 27--28, 2025.}
A data center cooling failure froze \$1 trillion in derivatives
trading for 11+ hours, affecting E-mini S\&P 500, Treasury futures,
crude oil, gold, and foreign exchange markets. A single point of
failure in physical infrastructure cascaded to global financial
markets \cite{cme_outage}.

\paragraph{FAA NOTAM system, January 11, 2023.}
Contract personnel accidentally deleted files while synchronizing
the primary and backup NOTAM databases. Because both systems
shared the same network infrastructure, corruption propagated
to the backup---a textbook atomicity failure where the backup
mechanism provided no fault isolation. Result: 1,300+ flight
cancellations, nearly 10,000 delays, and a 2-hour national
ground stop \cite{faa_notam}.

\subsection{The Economic Cost of FITO}

The cumulative evidence quantifies the economic burden of the
FITO category mistake:

\begin{itemize}
\item Gartner estimates average IT downtime cost at \$9,000 per
      minute (\$540,000/hour) as of 2024---a 61\% increase over
      the prior decade.
\item 44\% of organizations report hourly downtime costs exceeding
      \$1 million (Uptime Institute, 2025).
\item Single cloud outages produce losses in the hundreds of millions:
      \$500--650M (AWS October 2025), \$150M (AWS S3 2017),
      \$60M+ (Meta 2021).
\item AI training checkpoint overhead alone represents \$120--430M
      in wasted compute for frontier model runs approaching \$1B.
\item Financial markets lose approximately \$5 billion per year to
      latency arbitrage exploiting the gap between the NBBO
      simultaneity convention and physical causality
      \cite{borrill2026simultaneity}.
\end{itemize}

Every entry in this list traces to the same structural cause:
systems designed around the assumption that operations are
instantaneous state transitions, compensating for the failure
of that assumption with retry-and-recover protocols that
amplify the very failures they attempt to mask.

\begin{fitobox}[title=The Case for Transactional Integrity at the Network Layer]
The FITO category mistake is not a theoretical curiosity.
It is the structural root cause of cascading cloud outages,
database corruption, AI training waste, and financial system
failures totaling billions of dollars annually.

Eliminating retry amplification requires replacing the
unidirectional message---the FITO primitive---with a bilateral
transaction: an atomic interaction that completes for both
endpoints or fails for both. No messages ``in flight.''
No retry amplification.
\end{fitobox}


\begin{thebibliography}{40}

\bibitem{ferrite}
Bornholt et al,
``Specifying and Checking File System Crash Consistency Models,''
ASPLOS 2016.

\bibitem{ext4docs}
Linux Kernel Documentation,
``ext4 Filesystem Documentation,''
\url{https://docs.kernel.org/filesystems/ext4.html}

\bibitem{posix_rename}
POSIX,
``rename() Specification,''
\url{https://pubs.opengroup.org}

\bibitem{fastcommit}
Linux Kernel Documentation,
``ext4 Fast Commit,''
\url{https://lwn.net/Articles/842385}

\bibitem{vxfsintent}
VERITAS,
``VERITAS File System Administrator Guide.''

\bibitem{vxvmdrl}
VERITAS,
``VERITAS Volume Manager Administrator Guide.''

\bibitem{borrill2026icloud}
Borrill,
``Why iCloud Fails,''
arXiv:2602.19433

\bibitem{rebello2021}
Rebello et al,
``Can Applications Recover from fsync Failures?''
ACM Transactions on Storage, 2021.

\bibitem{nvme_cmdset_2024}
NVM Express, Inc.,
``NVM Express NVM Command Set Specification, Revision 1.1,''
(Revision ratified August 2024),
\url{https://nvmexpress.org/wp-content/uploads/NVM-Express-NVM-Command-Set-Specification-Revision-1.1-2024.08.05-Ratified.pdf}.

\bibitem{linux_writeback_cache_control}
Linux Kernel Documentation,
``Explicit volatile write back cache control,''
\url{https://docs.kernel.org/block/writeback_cache_control.html}.

\bibitem{nvme_flush_noop}
Discussion quoting NVMe specification behavior regarding Flush as a no-op when
volatile write cache is absent or disabled,
\url{https://news.ycombinator.com/item?id=46535896}.

\bibitem{ms_nvme_vwc_note}
Microsoft Documentation (NVMe VWC feature note),
``NVME\_CDW11\_FEATURE\_VOLATILE\_WRITE\_CACHE,''
\url{https://learn.microsoft.com/en-us/windows/win32/api/nvme/ns-nvme-nvme_cdw11_feature_volatile_write_cache}.

\bibitem{linux_queue_fua}
Linux Kernel Documentation (queue sysfs),
``queue-sysfs: fua,''
\url{https://www.infradead.org/~mchehab/kernel_docs/block/queue-sysfs.html}.

\bibitem{ssd_plp}
Kingston Technology,
``A Closer Look At SSD Power Loss Protection,''
\url{https://www.kingston.com/en/blog/servers-and-data-centers/ssd-power-loss-protection}.

\bibitem{herlihy1991}
M.~Herlihy,
``Wait-free synchronization,''
ACM Transactions on Programming Languages and Systems, 1991.

\bibitem{borrill2026impossibility}
P.~Borrill,
``The Impossibility of Atomicity Through Reads and Writes:
RDMA Semantics, Category Mistakes, and the Case for Swap-Based Architectures,''
DÆDÆLUS, 2026.

\bibitem{borrill2026rethinking}
P.~Borrill,
``Rethinking Atomicity: Counterfactual Transactions,''
DÆDÆLUS, 2026.

\bibitem{torvalds_nmi}
L.~Torvalds,
``Re: x86/entry: FRED CPU exception / NMI entry considerations,''
Linux Kernel Mailing List, 2023.
\url{https://lore.kernel.org/lkml/}

\bibitem{fred_intel}
Intel Corporation,
``Flexible Return and Event Delivery (FRED) Specification,''
2023.

\bibitem{desnoyers_rseq}
M.~Desnoyers,
``The 5-year journey to bring restartable sequences to Linux,''
EfficiOS Blog, February 2019.
\url{https://www.efficios.com/blog/2019/02/08/linux-restartable-sequences/}

\bibitem{mulliganstew2024}
P.~Borrill et al.,
``Mulligan Stew: Atomicity, Completeness, and
Reversible Protocols,''
Private discussion group on causal consistency and protocol-based
atomicity, 2024.
\url{https://mulligan-stew-portal.lovable.app}

\bibitem{borrill2026hermetic}
P.~Borrill,
``Deterministic Document Builds Under Eventually Consistent Storage,''
DÆDÆLUS, 2026.

% === Appendix references ===

\bibitem{gcp_june2025}
Google Cloud,
``Incident Report: Service Control Outage, June 12, 2025,''
\url{https://status.cloud.google.com/incidents/ow5i3PPK96RduMcb1SsW}.

\bibitem{aws_oct2025}
ThousandEyes,
``AWS Outage Analysis: October 20, 2025,''
\url{https://www.thousandeyes.com/blog/aws-outage-analysis-october-20-2025}.

\bibitem{aws_kinesis2020}
AWS,
``Summary of the Amazon Kinesis Event in the Northern Virginia (US-EAST-1) Region,''
November 25, 2020.

\bibitem{meta_oct2021}
Meta Engineering,
``More details about the October 4 outage,''
October 5, 2021.
\url{https://engineering.fb.com/2021/10/05/networking-traffic/outage-details/}.

\bibitem{cloudflare_sept2025}
Cloudflare,
``A deep dive into Cloudflare's September 12, 2025 dashboard and API outage,''
\url{https://blog.cloudflare.com/deep-dive-into-cloudflares-sept-12-dashboard-and-api-outage/}.

\bibitem{danluu_fsyncgate}
D.~Luu,
``Fsyncgate,''
\url{https://danluu.com/fsyncgate/}.

\bibitem{etcd_corruption}
etcd-io,
``etcd 3.5.0 panic: corrupted raft log,'' GitHub Issue \#13509,
2021.

\bibitem{mysql_doublewrite}
Percona,
``What is InnoDB Double Write and How Does It Work?''
\url{https://www.percona.com/blog/innodb-double-write/}.

\bibitem{nvme_plp_test}
R.~Bishop,
``I tested four NVMe SSDs from four vendors---half lose FLUSH'd data
on power loss,'' February 2022.

\bibitem{cern_corruption}
CERN,
``Data Corruption Research,'' 2007.
Reported in StorageMojo, September 2007.

\bibitem{netapp_corruption}
L.~Bairavasundaram et al.,
``An Analysis of Data Corruption in the Storage Stack,''
FAST 2008.

\bibitem{meta_sdc}
Meta Engineering,
``Silent Data Corruption at Scale,''
February 2021.
\url{https://engineering.fb.com/2021/02/23/data-infrastructure/silent-data-corruption/}.

\bibitem{ocp_sdc}
Open Compute Project,
``Silent Data Corruption in AI,'' OCP Whitepaper, 2024.

\bibitem{llama3_failures}
Meta,
``The Llama 3 Herd of Models,'' 2024.
Training infrastructure data: 419 interruptions across 54 days
on 16,384 H100 GPUs.

\bibitem{meta_mttf_study}
S.~Patel et al.,
``Revisiting Reliability in Large-Scale Machine Learning Research
Clusters,'' arXiv:2410.21680, 2024.

\bibitem{google_checkpoint}
Google Cloud Blog,
``Using multi-tier checkpointing for large AI training jobs,'' 2024.

\bibitem{aws_checkpoint}
AWS Blog,
``Architecting scalable checkpoint storage for large-scale ML
training on AWS,'' 2024.

\bibitem{knight_capital}
SEC,
``SEC Charges Knight Capital With Violations of Market Access Rule,''
Press Release 2013-222, October 16, 2013.

\bibitem{cme_outage}
CNBC,
``CME halts FX, commodities, futures trading after data center issue,''
November 28, 2025.

\bibitem{faa_notam}
US Department of Transportation,
``FAA NOTAM System Failure and Its Impacts,'' January 2023.

\bibitem{borrill2026simultaneity}
P.~Borrill,
``Engineered Simultaneity,''
DÆDÆLUS, 2026.

\end{thebibliography}
\end{document}